\documentclass[pre,aps,twocolumn]{revtex4}

\usepackage{graphicx}
\usepackage{amsmath,empheq}
\usepackage{amssymb}
\usepackage{ifthen}
\usepackage{booktabs}

\usepackage{graphicx}
\usepackage{dcolumn}
\usepackage{bm}
\usepackage{gensymb}
\usepackage{color}

\newcommand{\bb}{\begin{equation}}
\newcommand{\ee}{\end{equation}}
\newcommand{\ba}{\begin{eqnarray*}}
\newcommand{\ea}{\end{eqnarray*}}

\makeatletter
\def\@frameeq#1{%
  \framebox{$\,\displaystyle#1\hbox{\vrule height 2.4ex depth 1.4ex width 0pt}\,$}}
\newcommand\Equation[1]{$$\refstepcounter{equation}%
  \@frameeq{#1}%
  \eqno \hbox{\@eqnnum}$$\@ignoretrue\ignorespaces}
\newcommand\Displaystyle[1]{$$\@frameeq{#1}$$\@ignoretrue\ignorespaces}
\makeatother

\usepackage{longtable}

\newcounter{subfigcount}
\newcounter{figcount}
\newcommand{\subfloat}[3]{%
{\ifnum\thefigure=\thefigcount\stepcounter{subfigcount}%
\else\setcounter{figcount}{\thefigure}\setcounter{subfigcount}{1}\fi%
}%
\noindent%
\begin{minipage}[b]{#1}%
  \centering%
  {#3}\\[0pt]%
  (\alph{subfigcount})~#2%
\end{minipage}}%

\newcommand{\subfloatflex}[2]{%
{\ifnum\thefigure=\thefigcount\stepcounter{subfigcount}
\else\setcounter{figcount}{\thefigure}\setcounter{subfigcount}{1}\fi%
}%
\noindent%
\begin{minipage}[b]{\widthof{#2}}%
  \centering%
  {#2}\\[0pt]%
  (\alph{subfigcount})~#1%
\end{minipage}}%

\begin{document}

\normalsize

\title{Capillary condensation between non-parallel walls}

\author{Alexandr \surname{Malijevsk\'y}}
\affiliation{{Research group of Molecular Modelling, The Czech Academy of Sciences, Institute of Chemical Process Fundamentals, 165 02 Prague, Czech Republic;}
    {Department of Physical Chemistry, University of Chemical Technology Prague, 166 28 Prague, Czech Republic}}
\author{Ji\v r\'\i \hspace{0.001cm} \surname{Janek}}
\affiliation{{Department of Physical Chemistry, University of Chemical Technology Prague, 166 28 Prague, Czech Republic}}

\begin{abstract}
  \noindent We study the condensation of fluids confined by a pair of non-parallel plates of finite height $H$. We show that such a system experiences two types of condensation, termed single- and
  double-pinning, which can be characterized by one (single-pinning) or two (double-pinning) edge contact angles describing the shape of menisci pinned at the system edges. For both types of capillary
  condensation we formulate the Kelvin-like equation and determine the conditions under which the given type of condensation occurs. We construct the global phase diagram revealing a reentrant
  phenomenon pertinent to the change of the capillary condensation type upon varying the inclination of the walls. Asymptotic properties of the system are discussed and a link with related phase
  phenomena in different systems is made. Finally, we show that the change from a single- to a double-pinned state is a continuous transition, the character of which depends on the wetting properties
  of the walls.
\end{abstract}

\maketitle

\section{Introduction}

Confining fluids exhibiting liquid-gas phase separation below a critical temperature $T_c$ can profoundly change their phase behaviour \cite{RW,croxton,charvolin,henderson92,gelb}. If the confinement
is formed by a pair of parallel walls whose dimensions are macroscopic (effectively infinite) that are a distance $L$ apart,  the shift in the liquid-gas phase boundary from the vapour pressure
$p_{\rm sat}$ to $p_{\rm CC}(L)=p_{\rm sat}-\delta p_{\rm CC}(L)$, due to interplay between surface and finite-size effects, can be described by the classical Kelvin equation, according to which
 \bb
 \delta p_{\rm CC}(L)=\frac{2\gamma\cos\theta}{L}\,. \label{clas_kelvin}
 \ee
Here, $\gamma$ is the liquid-gas surface tension and  $\theta$ is Young's contact angle pertinent to a macroscopic liquid drop sitting on one of the isolated walls. This phenomenon of capillary
condensation, which persists up to the capillary critical point at a temperature $T_c(L)<T_c$ which is itself a subject of a finite-size change \cite{NF81,NF83}, has been a subject of numerous
theoretical studies, simulations and experiments, which revealed that the Kelvin equation remains quantitatively reliable even for slits several molecular diameters wide \cite{fisher,binder08,
  van,zhong,kim,yang,geim}.

\begin{figure*}[tbh!]
  \centering%
  \subfloat{0.99\columnwidth}{Single-pinned state.}{%
    \includegraphics[width=0.7\textwidth]{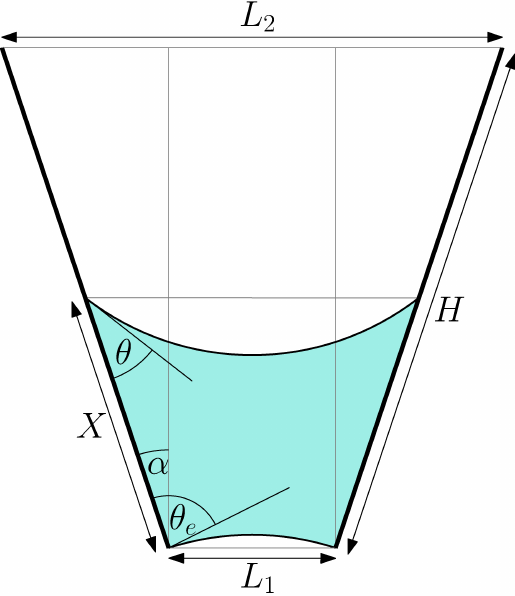}%
    \label{fig:single-pinned-scheme}%
  }%
  \hspace{0.02\columnwidth}%
  \subfloat{0.99\columnwidth}{Double-pinned state.}{%
    \includegraphics[width=0.7\textwidth]{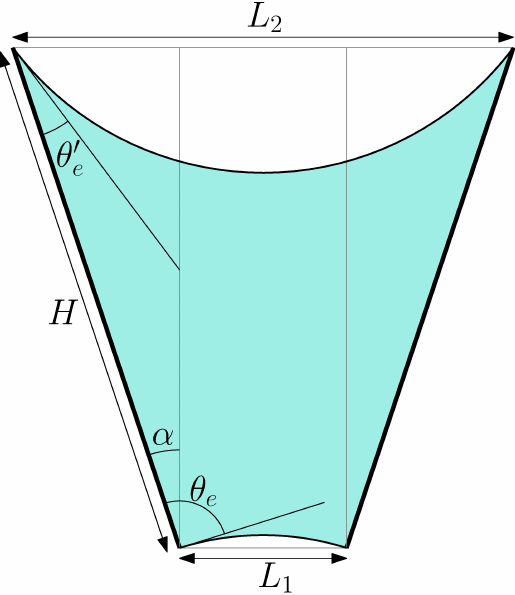}%
    \label{fig:double-pinned-scheme}%
  }%
  \caption{Schematic 2D illustration of two possible condensed states in a slit formed of two identical walls of finite length $H$, which both make an angle $\alpha$ with the vertical plane. The widths
    of the slit openings are $L_1$ and $L_2>L_1$ and the walls (with Young's contact angle $\theta$) are macroscopically deep. Panel a) refers to a single-pinned state which can be characterized
    by one edge contact angle $\theta_e$, whilst panel b) shows a double-pinned state characterized by two contact angles $\theta_e$ and $\theta_e'$.} \label{scheme}
\end{figure*}

However, the recent advances in nanoscale physics and the challenges associated with the fabrication of nano-devices called for further elucidation of phase phenomena induced by objects of
microscopic dimensions \cite{nano1}. In particular, a description of condensation in such systems requires a modification of the original Kelvin equation accounting properly for the effects at walls
boundaries and the system geometry. For instance, for plates of finite height $H$, but still parallel, it turns out that condensation appears always closer to saturation (i.e. at a higher pressure)
compared to infinite  ($H\to\infty$) walls and that the condensed state can be macroscopically characterized by an occurrence of two symmetric menisci that meet the ends of the walls at the
\emph{edge contact angle} $\theta_e$, which is always larger than the corresponding Young contact angle $\theta$ \cite{edge_slit}. This can be described using the modified Kelvin-like equation
\cite{edge_slit,depin21}
 \bb
 \delta p_{\rm CC}(L,H)=\frac{2\gamma\cos\theta_e}{L}\,, \label{dp_fs}
 \ee
determining the condensation pressure inside the finite slit $p_{\rm CC}(L,H)=p_{\rm sat}-\delta p_{\rm CC}(L,H)$, at which the edge contact angle satisfies
\bb
 \cos\theta_e=\cos\theta-\frac{L}{2H}\left[\sin\theta_e+\sec\theta_e\left(\frac{\pi}{2}-\theta_e\right)\right]\,. \label{thetae_fs}
 \ee

In this paper we show that the phase behaviour of confined fluids is significantly enriched if the confining walls are \emph{non-parallel} forming a \emph{truncated wedge} geometry. In this case, the
confinement (sketched in Fig.~\ref{scheme}) is formed of two identical plates of length $H$ that are both inclined (relative to the vertical, say) by an angle $\alpha$. The model capillary is thus of
finite height with $L_1$ and $L_2>L_1$ denoting the ``bottom'' and ``top'' opening widths. We assume that the depth of the capillary is macroscopic and that the system is translation-invariant along
the walls, implying that only one principal radius of curvature of the menisci is relevant. Hence, the model can be thought of either as an extension of that of the finite slit mentioned above or as
a part sliced off from a linear wedge with an opening angle $2\alpha$ \cite{hauge, rejmer, wood, bruschi, our_wedge}. Owing to the presence of two openings of different widths, the system allows for
condensation in two distinct ways: i) to a \emph{single-pinned} state (see Fig.~\ref{scheme}a), which is characterised by the presence of two menisci, such that one is pinned at the narrow slit end,
while the other is located inside the slit or ii) to a \emph{double-pinned} state  (see Fig.~\ref{scheme}b), in which case both menisci are located at the slit ends. It is then natural to ask, under
which conditions the given type of capillary condensation, if any, occurs. Associated with it, there are further natural questions. What are the analogues of the Kelvin equation for both types of
condensation? When does the system cease to condense? What is the role of the walls surface properties? What is the nature of the change from a single-pinned to a double-pinned state?

The remainder of the paper is organized as follows. In section II, we present a macroscopic theory of condensation, separately describing both possible types. Some of these results are then employed
in section III for an analysis of condensation inside a semi-infinite system \cite{terminology} where we primarily focus on the asymptotic behaviour of the system in the limit of $\alpha\to0$. The
main findings of our work are described in Section IV where we discuss the phenomenology of condensation  in the general case, construct the global phase diagram, determine the nature of the
depinning transition and present some analytic predictions that we compare with the numerical calculations. We conclude in section V with a detailed summary of the results.

\section{Two types of condensation} \label{theory}

In this section we formulate the Kelvin-like equations corresponding to both single-pinning and double-pinning condensation between non-parallel walls. To this end, we treat our capillary model as an
open system which is in equilibrium with a reservoir of the same subcritical temperature $T$ and the chemical potential $\mu$. Within this grand-canonical treatment an equilibrium state corresponds
to such a configuration, which minimizes the grand potential $\Omega$ which we will approximate using a simple macroscopic description. Clearly, the condition for condensation requires the value of
$\Omega$ of the ``empty'' (i.e., a gas-like) configuration and the pertinent condensed state, to be the same. Throughout the paper we will assume that the walls are ``hydrophilic'' with Young's
contact angle $\theta<\pi/2$. In the case of ``hydrophobic'' walls, $\theta>\pi/2$, capillary  condensation would be  replaced by capillary evaporation occurring at a pressure $p>p_{\rm sat}$, with
the role of gas and liquid phases being swapped. Some more details regarding the derivation of the Kelvin-like equations is given in Appendix \ref{ap_a}.

\subsection{Single pinning} \label{SP}
We start off with a description of condensation to a single-pinned state. The condensed system is characterized by the presence of two menisci, with one located  at the narrow end of the walls and
the other inside the capillary (see Fig.~\ref{scheme}a). Macroscopically, the shape of the menisci is determined by two characteristics: the Laplace radius of curvature $R$ and the angle at which the
menisci meet the walls. The Laplace radius, $R=\gamma/\delta p$, is of course the same for both menisci, where $\delta p=p_g-p_l$ is the pressure difference between the ambient gas and the
(metastable) condensed liquid phase. However,  while the  meniscus inside the capillary meets the walls with the equilibrium Young contact angle, $\theta$, the other meniscus which is pinned connects
the walls at the \emph{edge} contact angle, $\theta_e$. It is also clear, that while the location of the pinned meniscus is fixed, the one inside the capillary is free and its location $X$, varying
with $\delta p$, determines the portion of the capillary filled with liquid (cf. Fig.~\ref{scheme}a).  The macroscopic excess grand potential, per unit length, of the condensed system (relative to
the state where only gas is present), can be written as \bb \Omega^{\rm ex}_{\rm SP}=\delta pS+\gamma(\ell_1+\ell_2-2X\cos\theta) \label{om_sp}\,. \ee Here, the first term is the free-energy cost due
to the presence of the metastable liquid, the next two terms correspond to the surface free energies due to the presence of the menisci and the final term is the interfacial liquid-wall free energy
where Young's law has been used. Furthermore,
 \bb
 \ell_1=R(\pi+2\alpha-2\theta_e) \label{ell_left}\,,
  \ee
  and
  \bb
  \ell_2=R(\pi-2\alpha-2\theta) \label{ell_right}
  \ee
denote the arc-lengths of the pinned and the free menisci, respectively, and
 \bb
 S=S_{\rm tot}-S_1-S_2
 \ee
is the area corresponding to the region occupied by liquid. The latter was separated into the part $S_{\rm tot}$ of the trapezoid of length $X$,
 \bb
 S_{\rm tot}=\frac{4R^2\cos^2(\alpha+\theta)-L_1^2}{4}\cot\alpha\,,
 \ee
 and the parts
 \bb
 S_1=\frac{R^2(\pi+2\alpha-2\theta_e)-RL_1\sin(\theta_e-\alpha)}{2}
 \ee
 and
 \bb
 S_2=\frac{R^2(\pi-2\alpha-2\theta)}{2}-R^2\sin(\alpha+\theta)\cos(\alpha+\theta)\,,
 \ee
 corresponding to circular segments formed by the respective menisci. Finally, the length of the walls at the contact with liquid is
 \bb
 X=\frac{2R\cos(\alpha+\theta)-L_1}{2\sin\alpha}\,. \label{X}
 \ee

The location of the coexistence between the low-density state (filled only with gas) and the single-pinned state is determined from the condition $\Omega^{\rm ex}_{\rm SP}=0$. From this, it follows
that the single-pinned condensation occurs at the pressure $p_{\rm SP}=p_{\rm sat}-\delta p_{\rm SP}$, where
\bb
\delta p_{\rm SP}(L_1, H, \alpha)=\frac{2\gamma\cos(\theta_e-\alpha)}{L_1}\,, \label{ke_sp}
\ee
is the Kelvin-like equation for the condensation partial pressure with the edge contact angle $\theta_e$ given implicitly by the equation
\begin{widetext}
  \begin{equation}
    \cot\alpha\left[\cos^2(\alpha+\theta)-\cos^2(\alpha-\theta_e)\right]+\pi-\theta-\theta_e+
    \frac{\sin(2\alpha+2\theta)-\sin(2\alpha-2\theta_e)}{2}
    =\frac{2\cos\theta\left[\cos(\alpha+\theta)-\cos(\alpha-\theta_e)\right]}{\sin\alpha}\,. \label{ke_sp2}
  \end{equation}
\end{widetext}
Note that the location of the condensation can also be expressed in terms of the chemical potential $\mu_{\rm SP}=\mu_{\rm sat}(T)-\delta\mu_{\rm SP}$, where $\mu_{\rm sat}(T)$ is the value of the
chemical potential at saturation and $\delta p_{\rm SP}\approx\delta\mu_{\rm SP}(\rho_l-\rho_g)$ where $\rho_l$ and $\rho_g$ are the number densities of coexisting liquid and gas at the given
temperature \cite{marconi}.

\subsection{Double pinning} \label{DP}
Similar considerations lead to the Kelvin-like equation for double-pinned condensation, where, in contrast to the previous case, both menisci are pinned at the capillary ends, so that the walls are
at contact with liquid along their whole length (see Fig.~\ref{scheme}b). Now, the excess grand potential per unit length can be written as
\bb
\Omega^{\rm ex}_{\rm DP}=\delta pS+\gamma(\ell_1+\ell_2-2H\cos\theta)\,, \label{em_dp}
\ee
where the arc-length of the shorter meniscus, $\ell_1$, is given by (\ref{ell_left}), while the arc-length of the meniscus pinned with the
edge contact angle $\theta_e'(<\theta_e)$ at the wider end of the capillary is
\bb
\ell_2=R(\pi-2\theta_e'-2\alpha)\,. \label{ell_left2}
\ee

The area $S=S_{\rm tot}-\Delta S$ corresponding to the volume occupied by liquid is given by the total area between the walls, $S_{\rm tot}=H(L_1+L_2)\cos\alpha/2$,  reduced by the area of circular
segments due to the menisci
\bb
\Delta S=R^2(\pi-\theta_e-\theta_e')-\frac{R\left[L_1\sin(\theta_e-\alpha)+L_2\sin(\theta_e'+\alpha)\right]}{2}\,.
\ee
The edge contact angles are related to the Laplace radius of the menisci according to
\bb
R=\frac{L_1}{2\cos(\theta_e-\alpha)} \label{dp1}
\ee
and
\bb
R=\frac{L_2}{2\cos(\theta_e'+\alpha)}\,,\label{dp2}
\ee
which simply follows from the system geometry.

The phase boundary between low-density and double-pinned states occurs at the pressure $p_{\rm DP}=p_{\rm sat}-\delta p_{\rm DP}$, where
\bb
\delta p_{\rm DP}(L_1, H, \alpha)=\frac{\gamma}{R_{\rm DP}}\,. \label{ke_dp}
\ee
Here, the Laplace radius $R_{\rm DP}$ must, besides the geometric relations (\ref{dp1}) and (\ref{dp2}), satisfy the thermodynamic condition, $\Omega^{\rm ex}_{\rm DP}=0$, which yields
\begin{widetext}
  \begin{equation}
    \frac{L_2^2-L_1^2}{4}\cot\alpha+R^2_{\rm DP}(\pi-\theta_e-\theta_e')-\frac{(L_2-L_1)R_{\rm DP}\cos\theta}{\sin\alpha}
    +\frac{R_{\rm DP}}{2}\left[L_1\sin(\theta_e-\alpha)+L_2\sin(\theta_e'+\alpha)\right]=0\,. \label{dp3}
  \end{equation}
\end{widetext}
The phase boundary is thus determined by solving simultaneously  Eqs.~(\ref{dp1}), (\ref{dp2}), and (\ref{dp3}) for $\theta_e$, $\theta_e'$ and $R_{\rm DP}$; it is straightforward to verify that for
$\alpha=0$ Eqs.~(\ref{ke_dp}) and (\ref{dp3}) reduce to (\ref{dp_fs}) and (\ref{thetae_fs}), respectively.

\section{Condensation in semi-infinite slits} \label{semi-inf}

\begin{figure}[htb]
  \includegraphics[width=\columnwidth]{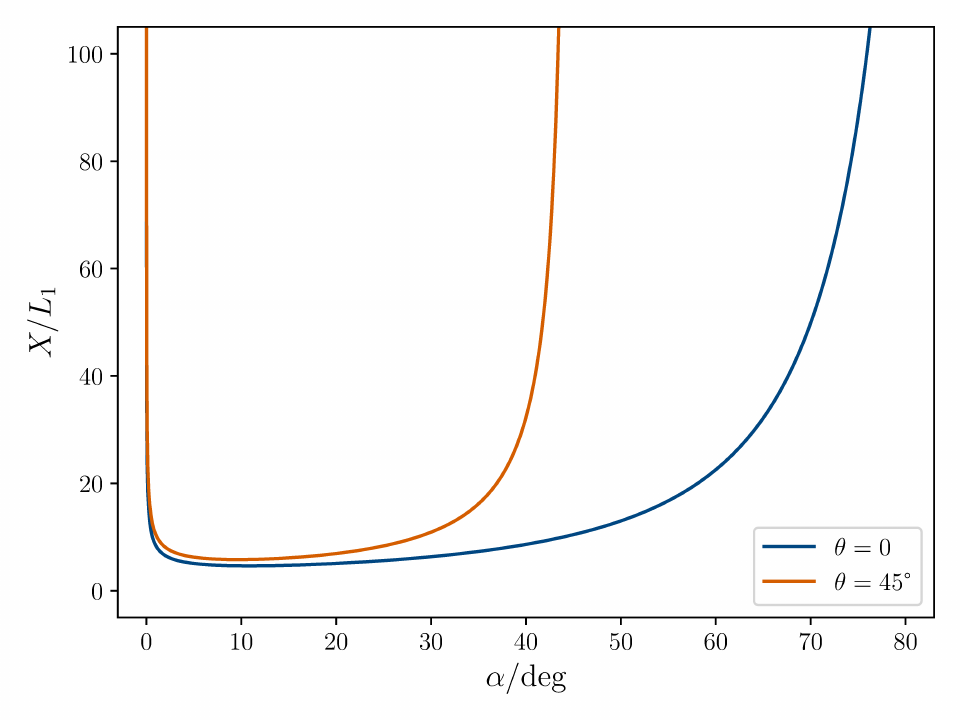}
  \caption{Illustration of a non-monotonic behaviour of the meniscus height $X$ at capillary condensation in a semi-infinite system as a function of the capillary opening angle $\alpha$; for $\theta=0$
    and $\theta=45\degree$. } \label{x-alpha_inf}
\end{figure}

Prior describing phase transitions in systems shown in Fig.~1, we begin with an analysis of semi-infinite systems that correspond to the limit of $H\to\infty$. In this case, which serves as a
pre-requisite to our main analysis, the situation is rather simple owing to the fact that only single-pinned condensation is possible. The condition under which the condensation may occur is
determined by an interplay between the thermodynamic angle $\theta$ and the geometric angle $\alpha$ (and does not depend on $L_1$, which is now the only macroscopic length scale). For $\alpha=0$,
the condensation is possible for any value of the contact angle $\theta\le\pi/2$ as in the case of a standard infinite slit  and the location of the meniscus, which forms at condensation, is
arbitrary \cite{semi_inf}. However, as soon as $\alpha>0$, the free-energy balance between the volume and interfacial contributions stabilizes the distance $X$ between the (lower) meniscus pinned at
the narrow end and the other one inside the capillary. This separation depends on $\alpha$ in a non-monotonic fashion, which is illustrated in Fig.~\ref{x-alpha_inf} and which has further
repercussions for finite capillaries as will be discussed later. Now for the given value of $\theta$, the maximum opening angle $\alpha_{\rm max}$ still allowing for condensation is the one, for
which the condensation occurs at the highest possible pressure below the bulk coexistence, $\delta p_{\rm SP}=0$, corresponding to flat menisci, implying that
\bb
\alpha_{\rm max}=\frac{\pi}{2}-\theta\,.   \label{alpha_max}
\ee
Alternatively, the maximum contact angle $\theta_{\rm max}$ allowing for condensation in the semi-infinite capillary with the fixed opening angle $\alpha$ is
\bb
\theta_{\rm max}=\frac{\pi}{2}-\alpha\,.   \label{theta_max}
\ee
The latter condition interpolates between that for capillary condensation in an infinite narrow slit ($\theta_{\rm max}=\pi/2$) and that for wetting transition on a planar wall ($\theta_{\rm
    max}=0$). Also note that it is just in the marginal case, $\alpha=\alpha_{\rm max}$, when $\theta_e=\theta$ and when {\emph{both}} menisci can freely move along the walls without any free energy
change.

\begin{figure}[htb]
  \includegraphics[width=\columnwidth]{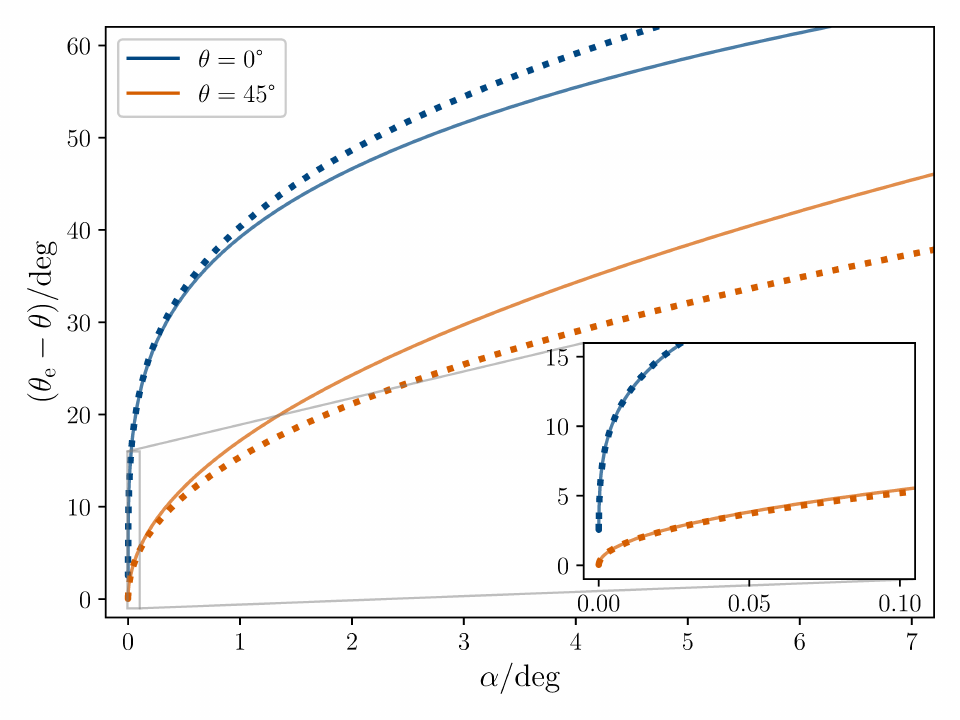}
  \caption{The behaviour of $\theta_e$ for small values of the opening angle $\alpha$ in semi-infinite slits for partially wet walls with $\theta=45$\degree{} and completely wet walls ($\theta=0$). The
    dotted lines refer to the numerical solution of Eq.~(\ref{ke_sp2}), while the solid lines represent the asymptotic results given by Eqs.~(\ref{thetae_as_pw}) and (\ref{thetae_as_cw}), respectively.}
  \label{thetae_as_inf}
\end{figure}

Next, we want to make a link between capillary condensation in our system and that in an infinite narrow slit. To this end, let us write $\Delta p_{\rm SP}(L_1,\infty, \alpha)=p_{\rm SP}(L_1,\infty,
  \alpha)-p_{\rm CC}(L_1)$ for the pressure difference between the respective condensation pressures. The difference is positive for any $\alpha>0$ and we wish to know how it vanishes in the limit of
narrowing the capillary. Here, we have to distinguish between partially and completely wet walls. For partially wet walls, $\theta>0$, we find that to lowest order in $\alpha$, the edge contact angle
approaches Young's contact angle according to
\bb
\theta_e=\theta+c(\theta)\sqrt{\alpha}\,,\;\;\;{\rm as}\; \alpha\to0  \label{thetae_as_pw}
\ee
with the amplitude $c(\theta)=\sqrt{(\pi-2\theta+\sin(2\theta)}/\sin\theta$ and which implies that
\bb
\Delta p_{\rm SP}(L_1,\infty, \alpha)=\frac{2\gamma c(\theta)\sin\theta\sqrt{\alpha}}{L_1}+{\cal{O}}(\alpha)\,.
\ee

This contrasts with the complete wetting regime, $\theta=0$, for which we obtain
\bb
\theta_e=\theta+(4\pi\alpha)^{\frac{1}{4}}\,,\;\;\;{\rm as}\; \alpha\to0 \label{thetae_as_cw}
\ee
and
\bb
\Delta p_{\rm SP}(L_1,\infty, \alpha)=\frac{2\gamma\sin\theta(4\pi\alpha)^{\frac{1}{4}}}{L_1}+{\cal{O}}(\sqrt{\alpha})\,.
\ee
The test of the asymptotic relations (\ref{thetae_as_pw}) and (\ref{thetae_as_cw}) against the numerical results obtained from  Eq.~(\ref{ke_sp}) for both partial and complete wetting is shown in
Fig.~\ref{thetae_as_inf}.

\section{Condensation in finite slits}

\subsection{Phase diagrams}

\begin{figure*}[ht]
  \subfloat{\columnwidth}{}{%
    \includegraphics[width=\textwidth]{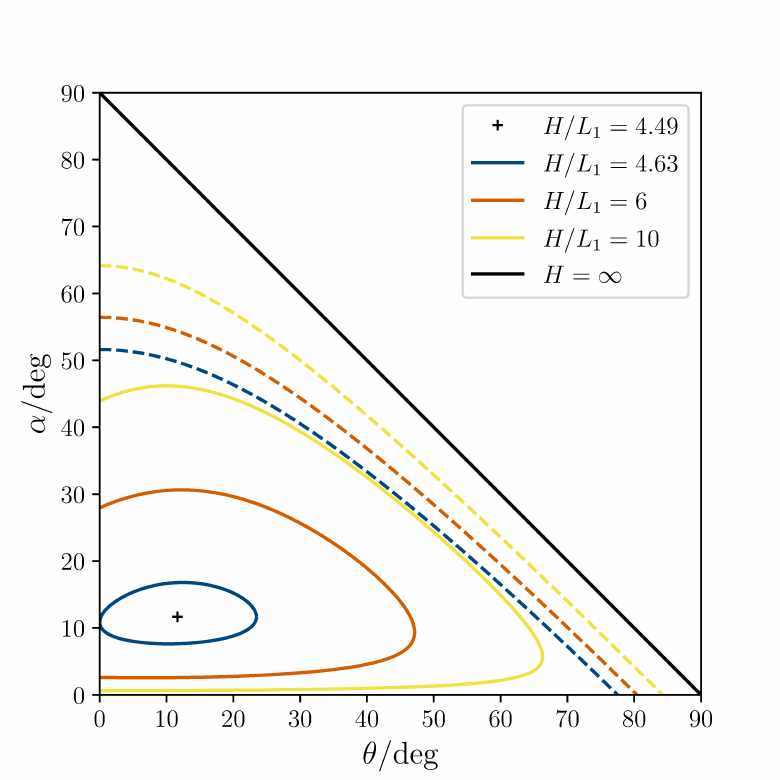}%
    \label{fig:phase_diag-a}%
  }%
  \hspace{\columnsep}%
  \subfloat{\columnwidth}{}{%
    \includegraphics*[width=\textwidth]{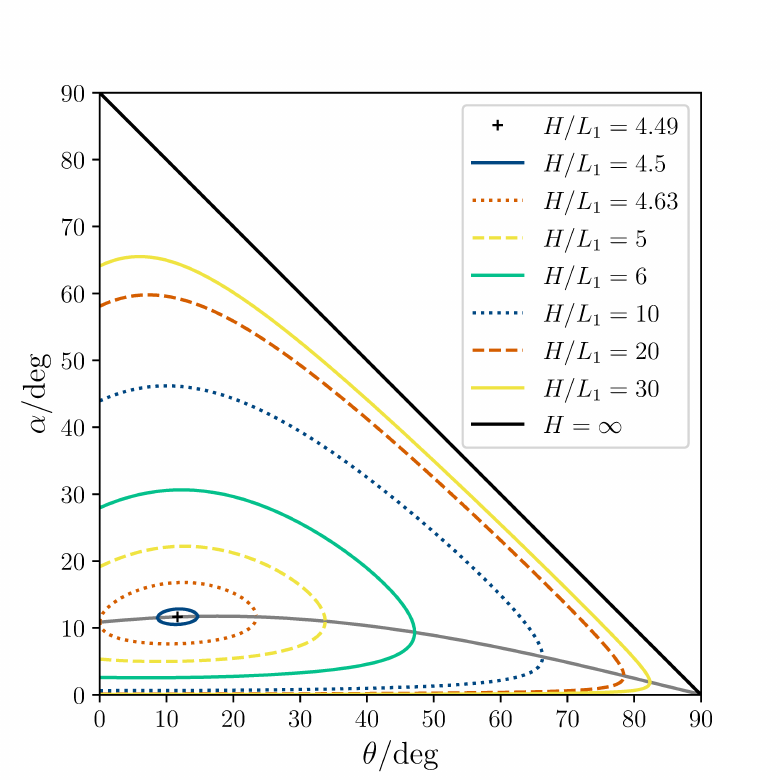}%
    \label{fig:phase_diag-b}%
  }%
  \caption{a) Global phase diagram specifying phase boundaries displayed in the $\alpha$-$\theta$ projection for several values of $H$. The areas bounded by the solid curves referring to a given value
    of $H$ correspond to a set of parameters, for which the system condenses to single-pinned states. The area outside the given solid curve and below the pertinent dashed line corresponds to a set of
    parameters, for which the system condenses to a double-pinned state. Above the dashed lines, representing the curves given by Eq.~(\ref{alpha_max2}), the opening angle is too large for the system to
    condense; these curves tend in the limit of $H\to\infty$ to the solid black line $\pi/2-\theta$ (cf. Eq.~(\ref{alpha_max2})). b) Same as a) except that in place of the lines showing the condensation
    limit,  the line connecting the ``turning points'' corresponding to $X_{\rm min}(\theta)$ shown in Fig.~\ref{xmin} is displayed.} \label{phase_diag}
\end{figure*}

\begin{figure}[hbt]
  \includegraphics[width=\columnwidth]{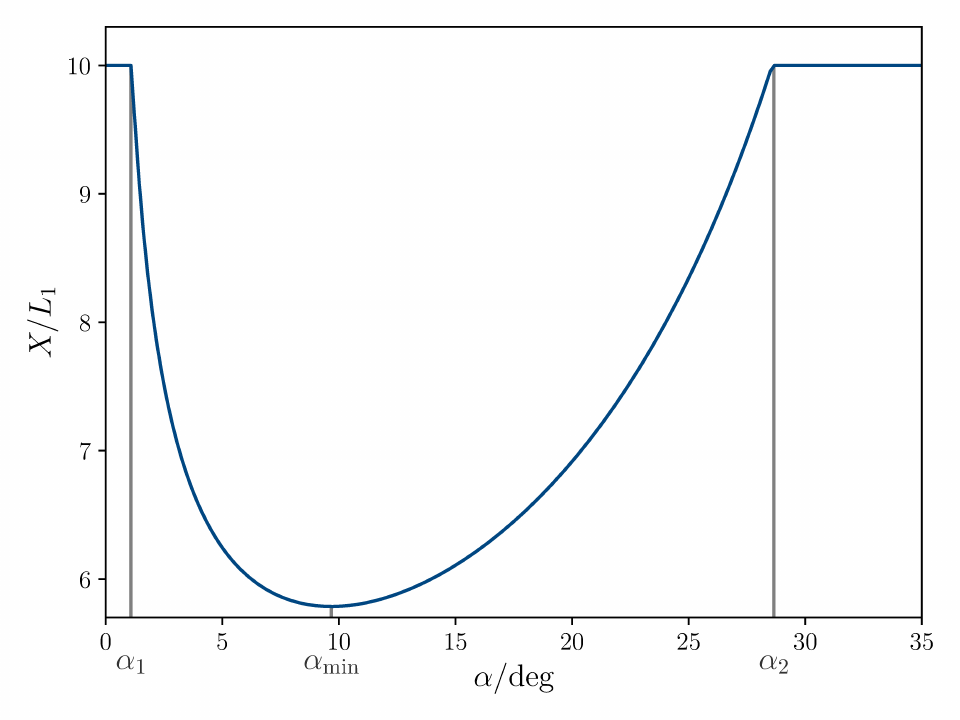}
  \caption{The meniscus height $X$ at capillary condensation as a function of the opening angle $\alpha$ for $H/L_1=10$ and $\theta=45\degree$. For $\alpha<\alpha_1$ and $\alpha>\alpha_2$,
  $X=H$, meaning that the system is in a double-pinned state. For $\alpha_1<\alpha<\alpha_2$, the system condenses to a single-pinned state and $X(\alpha)$ is non-monotonic with a minimum  at
  $\alpha=\alpha_{\rm min}\approx9.7$\degree, at which $X=X_{\rm min}\approx5.8\,L_1$ (cf. Fig.~\ref{xmin}). } \label{x}
\end{figure}

\begin{figure}[t]
  \includegraphics[width=\columnwidth]{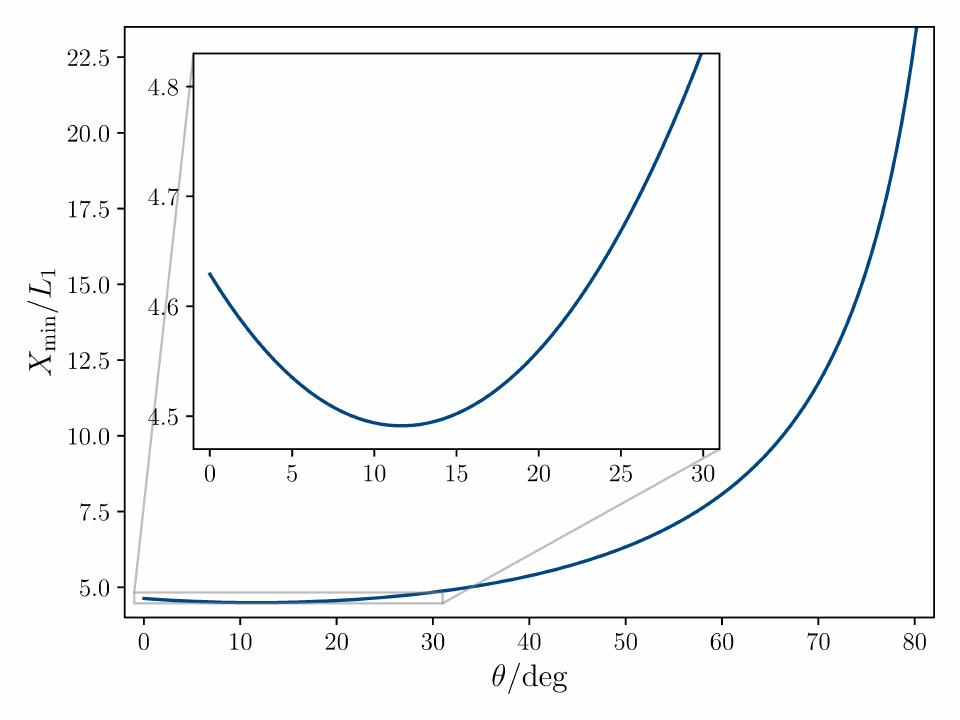}
  \caption{The dependence of the minimal meniscus height $X$ of a condensed, single-pinned state on $\theta$. } \label{xmin}
\end{figure}

We now turn to non-parallel slits formed of walls of finite length $H$. In contrast to the semi-infinite case, there are now two types of condensation and we wish to specify conditions determining
their realization.  These will be expressed in the parameter space of $H$, $\alpha$ and $\theta$, with  $L_2=L_1+2H\sin\alpha$ and $L_1$ taken as a unit of length. Prior to it, however,  we start
again with a formulation of requirements that the system must obey to experience capillary condensation at all, expressed in terms of the geometrical parameters $\alpha$ and $H$, for a given value of
$\theta$. The marginal conditions allowing for capillary condensation follow from the condition $\Delta\Omega^{\rm ex}_{\rm DP}=0$ at the maximal pressure, $p=p_{\rm sat}$, i.e. when the menisci are
flat. This implies that the maximal value $\alpha_{\rm max}$ of the opening angle is (cf. Appendix \ref{ap_b})
 \bb
 \sin\alpha_{\rm max}=\cos\theta-\frac{L_1}{H}\,,  \label{alpha_max2}
 \ee
 with $H\ge L_1\sec\theta$. For $H<L_1\sec\theta$ no condensation is possible for any value of $\alpha$.

Note that Eq.~(\ref{alpha_max2}) is a generalization of the condition (\ref{alpha_max}) for the semi-infinite case and is more stringent. The presence of the last term in Eq.~(\ref{alpha_max2})
implies that along the saturation path $p=p_{\rm sat}(T)$ the system will no longer condense at the wedge-filling phase boundary but at a temperature $T_f(\alpha)<T(\alpha,H)<T_w$.

\begin{figure*}[bht]
  \subfloatflex{0\degree}{%
    \includegraphics[scale=0.69]{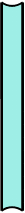}\hspace*{4.5ex}%
    \label{fig:state-alpha-0}%
  }\hspace{0.4em}%
  \subfloatflex{1\degree}{%
    \hspace*{4.5ex}\includegraphics[scale=0.69]{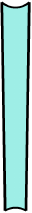}\hspace*{4.5ex}%
    \label{fig:state-alpha-1}%
  }\hspace{0.4em}%
  \subfloatflex{1.5\degree}{%
    \hspace*{4.5ex}\includegraphics[scale=0.69]{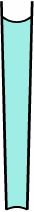}\hspace*{4.5ex}%
    \label{fig:state-alpha-1-5}%
  }\hspace{0.4em}%
  \subfloatflex{5\degree}{%
    \hspace*{4ex}\includegraphics[scale=0.69]{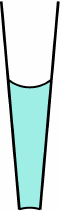}\hspace*{4ex}%
    \label{fig:state-alpha-5}%
  }\hspace{0.7em}%
  \subfloatflex{10\degree}{%
    \hspace*{2ex}\includegraphics[scale=0.69]{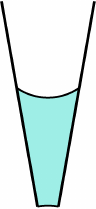}\hspace*{2ex}%
    \label{fig:state-alpha-10}%
  }\hspace{0.7em}%
  \subfloatflex{20\degree}{%
    \includegraphics[scale=0.69]{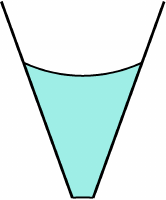}%
    \label{fig:state-alpha-20}%
  }\hspace{0.7em}%
  \subfloatflex{29\degree}{%
    \includegraphics[scale=0.69]{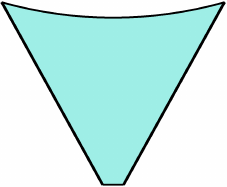}%
    \label{fig:state-alpha-29}%
  }\hspace{0.7em}%
  \subfloatflex{37\degree}{%
    \includegraphics[scale=0.69]{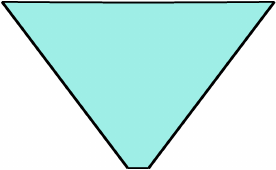}%
    \label{fig:state-alpha-37}%
  }%
  \caption{A sequence illustrating capillary condensed states of a system with $H=10\,L_1$ and $\theta=45$\degree{} for increasing $\alpha$. } \label{states}
\end{figure*}

Alternatively, Eq.~(\ref{alpha_max2}) can be recast to the form determining the minimal length of the walls, $H_{\rm min}$, required for the presence of capillary condensation for the given value of
the opening angle $\alpha<\alpha_{\rm max}$:
\bb
H_{\rm min }=\frac{L_1}{\cos\theta-\sin\alpha}\,.  \label{H_min}
\ee
We recall that these conditions were obtained by assuming that condensation at $p=p_{\rm sat}$ is of a double-pinned type. This must be the case, since condensation to a single-pinned state at
$p_{\rm sat}$ would require that $\sin\alpha<\cos\theta$ (in order $\Omega^{\rm ex}_{\rm SP}<\Omega^{\rm ex}_{\rm DP}$), in which case, however, $\Omega^{\rm ex}_{\rm SP}>0$ precluding the
condensation.

The main  results summarising the macroscopic predictions from section \ref{theory} are shown in Fig.~\ref{phase_diag}. Here, the global phase diagram in the  $\alpha$-$\theta$ plane is presented for
several representative values of $H$.  Fig.~\ref{phase_diag}a specifies, for each ratio of $H/L_1$, the values of $\alpha$ and $\theta$ corresponding (same colour) to the given type of capillary
condensation. The regions bounded by the solid lines correspond to the parameters, for which the system condenses to single-pinned states. The area outside these regions but below the corresponding
dashed line refers to double-pinning; the dashed lines show the maximal opening angle $\alpha_{\rm max}(\theta)$ above which no condensation is possible. Note that these lines straighten up and shift
to higher and higher values of $\theta$ as $H$ increases, approaching the line $\alpha=\pi-\theta$ pertinent to the semi-infinite case, which is also displayed.

These results reveal two noteworthy features which we want to point out:

\emph{Firstly}, the condensation exhibits a re-entrant phenomenon along the path of constant $\theta$ on varying the opening angle $\alpha$. This follows from the very fact that the boundaries of
this path always correspond to double-pinned states: the case  $\alpha=\alpha_{\rm max}$ was discussed above, while for $\alpha=0$ only double-pinned states are accessible \cite{edge_slit}. Hence,
within the interval of $\theta(H)$ allowing for single-pinned states, the increase of $\alpha$ will first switch from double- to single-pinning before turning back to double-pinning condensation.

The change in condensation type occurs for the values of $\alpha$, solving the equation
\begin{widetext}
  \bb
  c^2\cot\alpha(1-q^2)+\pi-\theta-\cos^{-1}(qc)-\alpha+c\sqrt{1-c^2}+qc\sqrt{1-q^2c^2}=\frac{2c(1-q)\cos\theta}{\sin\alpha}\,,       \label{triple}
  \ee
\end{widetext}
where we used the abbreviations $q=L_1/L_2$ and $c=\cos(\theta+\alpha)$ (see Appendix \ref{ap_c} for more details). Eq.~(\ref{triple}) determines the parameters for which both types of condensation
coincide, i.e., when $\theta_e'=\theta$, or equivalently $X=H$. This can be illustrated in Fig.~\ref{x} where a typical dependence of $X(\alpha)$ is shown. As for the semi-infinite case, the
dependence is non-monotonic but now, of course, is bounded by the length of the walls. The highlighted values of the opening angle $\alpha_1$ and $\alpha_2$ are those for which
$X(\alpha_1)=X(\alpha_2)=H$, both solving Eq.~(\ref{triple}).

The third significant value of the opening angle is $\alpha_{\rm min}$, for which $X(\alpha)$ reaches its minimum. This value can be determined analytically by solving $\partial X/\partial \alpha=0$,
using
\bb
\left(\frac{\partial R}{\partial \alpha}\right)_\theta=-\frac{\left(\frac{\partial\Omega^{\rm ex}_{\rm SP}}{\partial\alpha}\right)_\theta}{\left(\frac{\partial\Omega^{\rm ex}_{\rm SP}}{\partial
    \theta_e}\right)_\theta}\cdot\left(\frac{\partial R}{\partial\theta_e}\right)_\alpha\,,  \label{implicit}
\ee
where $(\partial R/\partial\theta_e)_\alpha=L_1\sec(\theta_e-\alpha)\tan(\theta_e-\alpha)/2$, as follows from Eq.~({\ref{ke_sp}). The dependence of $X_{\rm min}$ on $\theta$ is shown in
Fig.~\ref{xmin} and can be interpreted as the minimal length of walls $H$ allowing, for the given contact angle $\theta$, condensation to a single-pinned state. The dependence corresponds to the line
shown in Fig.~\ref{phase_diag}b connecting the ``turning points'' of the lines bounding the regions of single-pinning condensation, which are the single solutions of Eq.~(\ref{triple}). Note that the
graph $X_{\rm min}(\theta)$ itself exhibits a minimum for $\theta\approx 11.65\degree$ for which $X_{\rm min}\approx4.5\,L_1$.

\emph{Secondly}, the lines specifying the region of single-pinned condensation always possess a ``turning point'' at $\theta^+(H)$, which can be interpreted as the contact angle above which only
double-pinning is possible for arbitrary $\alpha$ (for the given $H$). However, there exists a specific value of the system size $\tilde{H}\approx4.63\,L_1$, such that for $H<\tilde{H}$ the lines
enclosing single-pinned condensation exhibit two such points --  $\theta^+(H)$ and  $\theta^-(H)$ -- meaning that single-pinning is only possible within the interval $\theta^-(H)<\theta<\theta^+(H)$.
As $H$ further decreases, the single-pinning region shrinks and ultimately vanishes when $H=H_c\approx 4.5\,L_1$, which corresponds to the minimum of $X_{\rm min}(\theta)$ shown in Fig.~\ref{xmin}.

We briefly summarise the present results. At first, our system formed of a pair of walls of length $H$ with Young's contact angle $\theta$ may only ever experience capillary condensation provided the
walls are long enough, such that $H\ge L_1\sec\theta$. Under these conditions two types of capillary condensation are in principle possible: i) to a single-pinned state characterized by an edge
contact angle $\theta_e$ at the pressure given by the generalized Kelvin equation (\ref{ke_sp}) or, more commonly, ii) to a double-pinned state characterized by two edge contact angles $\theta_e$ and
$\theta_e'$ at the pressure given by the generalized Kelvin equation (\ref{ke_dp}). The latter type always takes place for acute angles when the walls are almost parallel; however, as $\alpha$
increases the upper edge contact angle $\theta_e'$ decreases and eventually depins from the top edges when $\theta_e'=\theta$, in which case the bottom edge contact angle $\theta_e$ takes the value
given by Eq.~(\ref{triple}). Under further widening of the system, the condensation will lead to a single-pinned state, so that a portion of the system filled with liquid (proportional to $X$)
decreases rapidly with $\alpha$. This decrease continues up to the point given by the solution of Eq.~(\ref{implicit}), at which the upper meniscus reaches its minimal position, $X_{\rm min}$, and
beyond which the meniscus location approaches the upper edges again. The re-entrance to the double-pinned condensation is accomplished at such a value of $\alpha$ that corresponds to the second
solution of Eq.~(\ref{triple}) (when $X=H$).   The condensation to a double-pinned state persists up to $\alpha_{\rm max}$ as given by Eq.~(\ref{alpha_max2}), which generalizes the condition for the
filling transition in a linear wedge. For $\alpha>\alpha_{\rm max}$, the free-energy cost for the presence of the metastable liquid becomes too high for the system to condense. This phenomenology is
illustrated in Fig.~\ref{states}, where a sequence of equilibrium configurations of a system with $H=10\,L_1$ and $\theta=45$\degree{} is depicted for several values of $\alpha$. A corresponding
movie is shown in the Supplementary Material \cite{SM}.

\subsection{Asymptotic behaviour of capillary condensation}

\begin{figure*}[bpht]
  \centering%
  \subfloat{0.32\textwidth}{}{%
    \includegraphics[width=\textwidth]{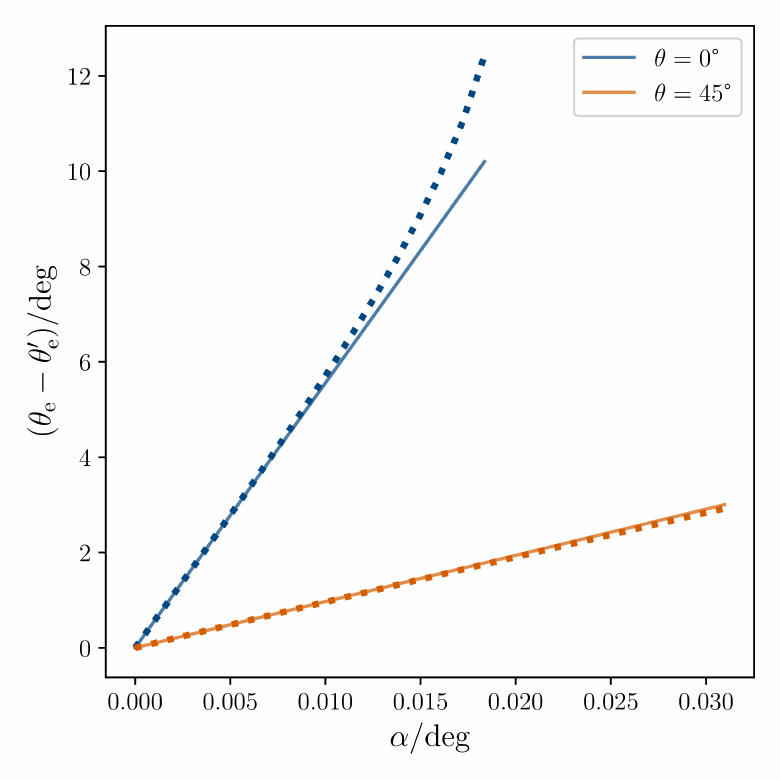}%
    \label{fig:alpha0-31}%
  }%
  \hspace{0.01\textwidth}%
  \subfloat{0.32\textwidth}{}{%
    \includegraphics[width=\textwidth]{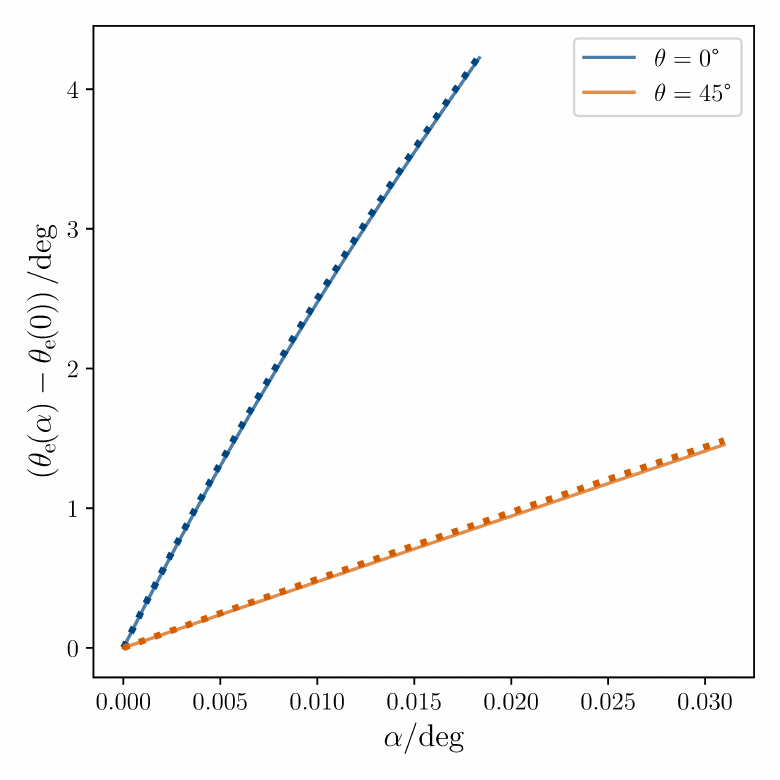}%
    \label{fig:alpha0-costhetae}%
  }%
  \hspace{0.01\textwidth}%
  \subfloat{0.32\textwidth}{}{%
    \includegraphics[width=\textwidth]{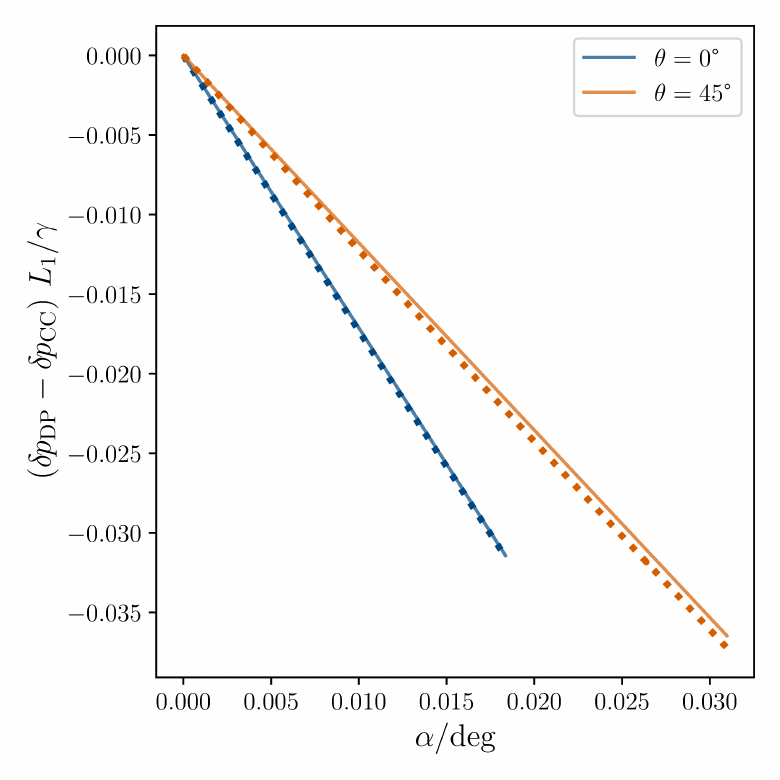}%
    \label{fig:alpha0-dpdp}%
  }%
  \caption{The numerical test of the asymptotic results given by a) Eq.~(\ref{dthetae}), b) Eq.~(\ref{costheta_e_alpha}), and c) Eq.~(\ref{p_alpha0}). The analytic predictions (solid lines) are compared with the numerical
    solutions of Eqs.~(\ref{ke_dp})--(\ref{dp3}) (symbols)  for partially wet walls ($\theta=45$\degree) and completely wet walls ($\theta=0$\degree) with $H=20\,L_1$. } \label{alpha0}
\end{figure*}

In this section, we analyze asymptotic features of capillary condensation, starting with its behaviour in the limit of $\alpha\to0$. This, for any finite value of $H$, contrasts with the case of
semi-infinite capillaries examined in section \ref{semi-inf}, since now the relevant type of condensation is double-pinning. From Eqs.~(\ref{dp1}) and (\ref{dp2}) it follows that to first order in
$\alpha$ the relation between the edge contact angles is
\bb
\theta_e-\theta_e'\approx 2\alpha\left(1+\frac{\cot\theta_e}{a}\right)\,, \label{dthetae}
\ee
where we have temporarily introduced the abbreviation, $a=L_1/H$. To same order, the corresponding Laplace radius is
\bb
R\approx\frac{L_1}{2}\sec\theta_e\left(1-\alpha\tan\theta_e\right)\,.
\ee
In what follows, we will further assume that the length-to-width ratio of the slits is large and will drop the terms of the order of $a\alpha$. The free energy balance, Eq.~(\ref{dp3}),
can then be written as
\begin{eqnarray}
  &&\alpha+a+\frac{a^2}{4}\sec^2\theta_e(\pi-2\theta_2)-a\cos\theta\sec\theta_e\nonumber\\
  &&+\frac{a^2}{2}\tan\theta_e+{\cal{O}}(\alpha^2,a\alpha)=0\,, \label{theta_e_alpha}
\end{eqnarray}
which yields the relation between $\theta_e(\alpha)$ and the edge contact angle for parallel slits, $\theta_e(0)$,
\bb
\cos\theta_e(\alpha)=\cos\theta_e(0)\left(1-\frac{\alpha H}{L_1}\right)\,,\;\;\;(\alpha\to0)\,. \label{costheta_e_alpha}
\ee
This implies that $\theta_e(\alpha)$ decays to $\theta_e(0)$ linearly, and, in contrast to the semi-infinite case, the asymptotic behaviour is the same both for partially and completely wet walls.
Furthermore, we can also write for the shift in the corresponding partial pressures:
\bb
\delta p_{\rm DP}=\delta p_{\rm CC}(L_1,H)\left[1-\left(\frac{H}{L_1}-\tan\theta_e(0)\right)\alpha\right]\,,\;\;\;(\alpha\to0)\,. \label{p_alpha0}
\ee
These asymptotic relations are confirmed  by comparing with the numerical results obtained from Eqs.~(\ref{ke_dp})--(\ref{dp3}) as shown in Fig.~\ref{alpha0}.

\begin{figure}[bthp]
  \includegraphics[width=\columnwidth]{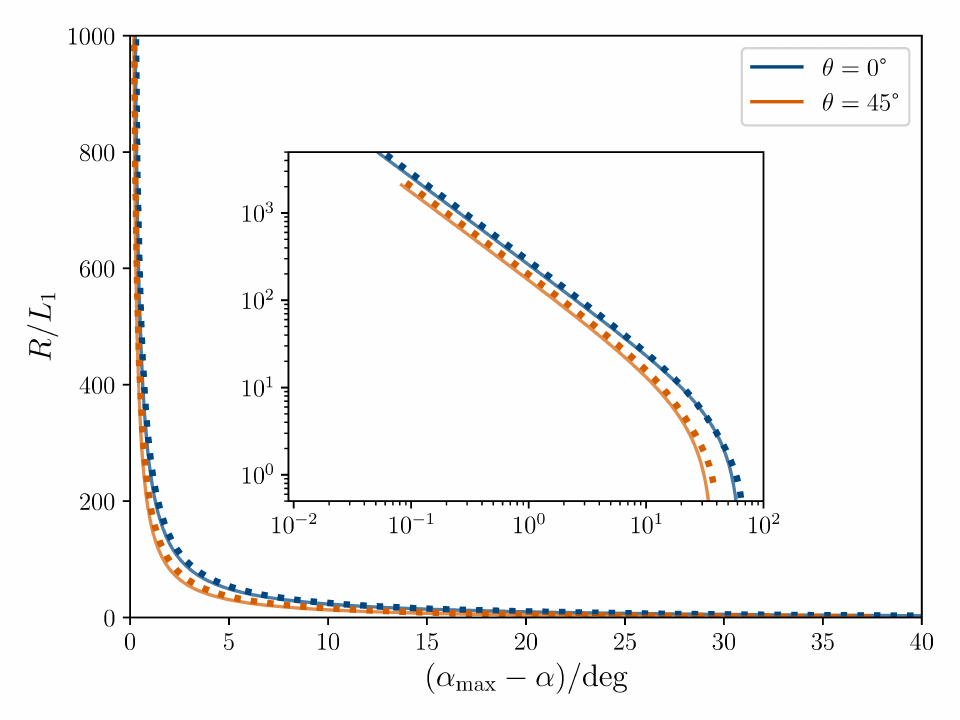}
  \caption{The numerical test of the power law (\ref{asym_R}) showing the divergence of the Laplace radius of the menisci upon widening the pore up to the condensation limit $\alpha_{\rm max}$. The
    analytic predictions (solid lines) are compared with the numerical solutions of Eqs.~(\ref{ke_dp})--(\ref{dp3}) (dotted) for systems with $H=10\,L_1$ and $\theta=0$ and 45\degree.} \label{divR}
\end{figure}

Next, we wish to investigate the growth of the equilibrium Laplace radius $R$ of the menisci along the double-pinned phase boundary, as the maximal opening angle $\alpha_{\rm max}$  is approached.
From Eqs.~(\ref{dp1}) and (\ref{dp2})  it follows that as $\alpha\to\alpha_{\rm max}$ (and $R\to\infty$), $\theta_e$ and $\theta_e'$ tend to adopt their respective asymptotic values, $\pi/2+\alpha$
and $\pi/2-\alpha$, according to \bb
\theta_e\sim\frac{\pi}{2}+\alpha-\frac{L_1}{2R}\,,\;\;\;\theta_e'\sim\frac{\pi}{2}-\alpha-\frac{L_2}{2R}\,.  \label{theta_asym}
\ee
In this limit, Eq.~(\ref{dp3}) reduces, upon substituting from (\ref{theta_asym}), to
\bb
L_1+L_2+\frac{(L_1-L_2)\cos\theta}{\sin\alpha}+\frac{(L_2^2-L_1^2)\cot\alpha}{4R}=0\,,
\ee
bearing in mind that $L_2$ itself is a function of $\alpha$. This implies the power law for the growth of the menisci
\bb
R\sim\frac{H\sin\alpha}{2}(\alpha_{\rm max}-\alpha)^{-1}\,,\;\;\;(\alpha\to\alpha_{\rm max})\,, \label{asym_R}
\ee
which is verified by a comparison with the exact numerical solution of Eqs.~(\ref{dp1})--(\ref{dp3}) as shown in Fig.~\ref{divR}.

\subsection{Depinning transition}

\begin{figure*}
  \subfloat{\columnwidth}{}{%
    \includegraphics[width=\textwidth]{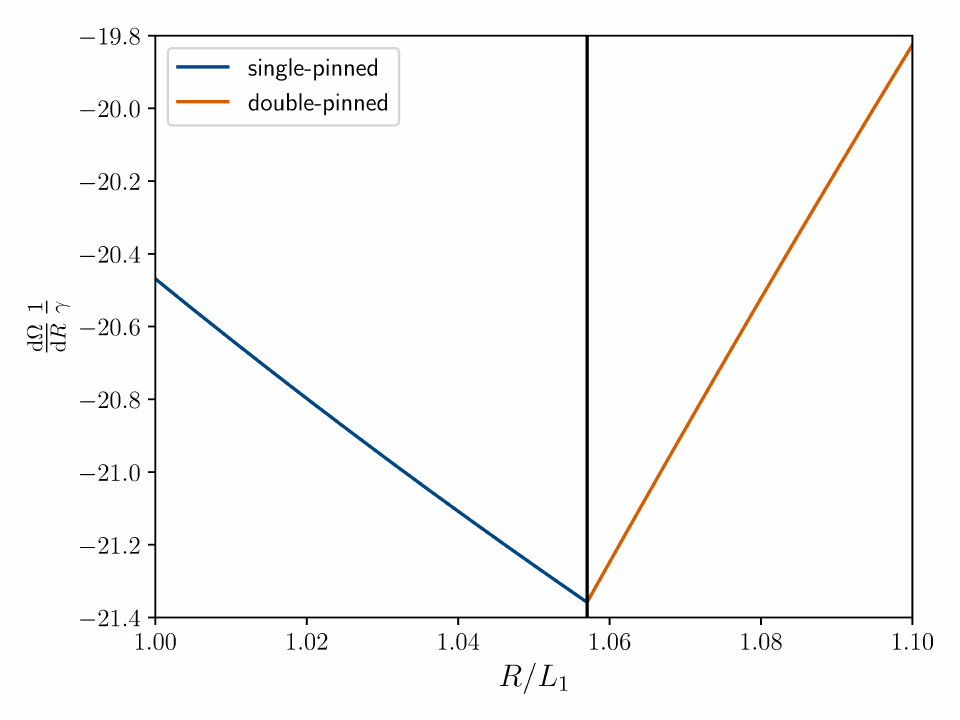}%
    \label{fig:domega_pw}%
  }%
  \subfloat{\columnwidth}{}{%
    \includegraphics[width=\textwidth]{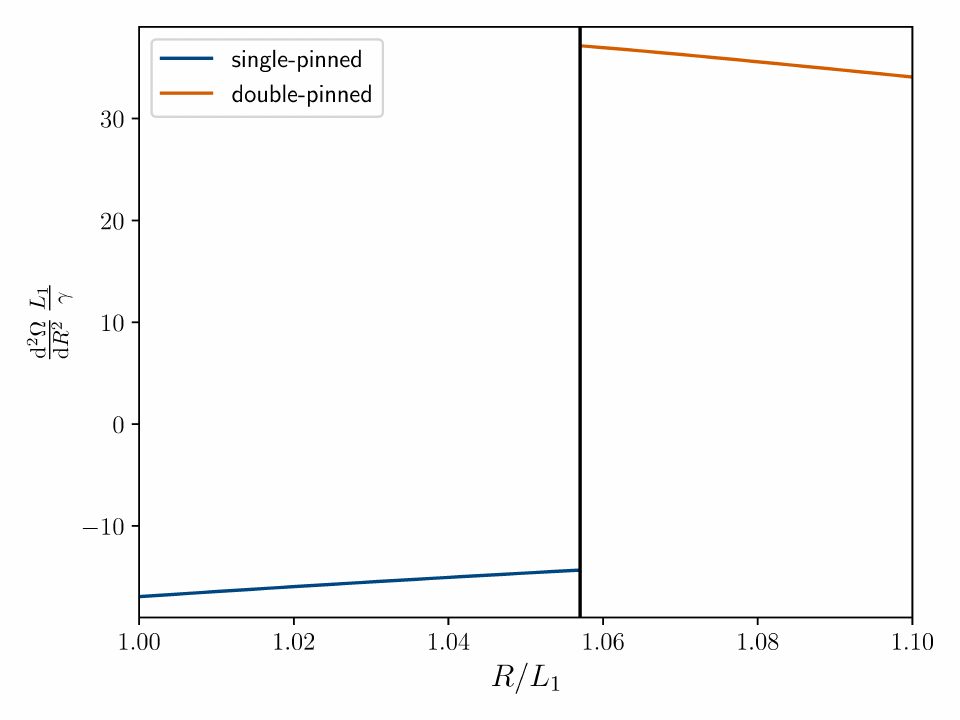}%
    \label{fig:ddomega_pw}%
  }%
  \caption{The illustrative behaviour of first (a) and second (b) derivatives of the grand potential of the single-pinned phase (for $R<R_{\rm depin}$) and the double-pinned phase (for $R>R_{\rm
          depin}$) where  $R_{\rm depin}\equiv\gamma/\delta p_{\rm depin}$ (denoted by the vertical line) corresponds to the depinning transition of the system formed of partially wet walls with $\theta=0.3$.
    The other parameters of the system are $L_2=2\,L_1$ and $\alpha=0.03$. }\label{domega_pw}
\end{figure*}

\begin{figure}
  \subfloat{\columnwidth}{}{%
    \includegraphics[width=\textwidth]{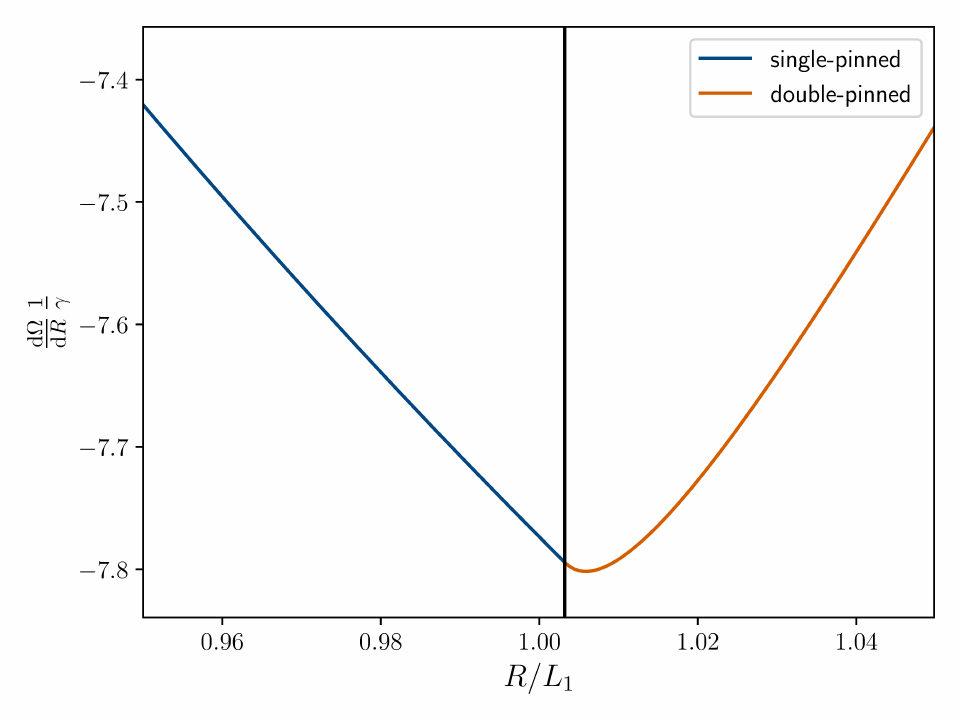}%
    \label{fig:domega_cw}%
  }%
  \\[0pt]%
  \subfloat{\columnwidth}{}{%
    \includegraphics[width=\textwidth]{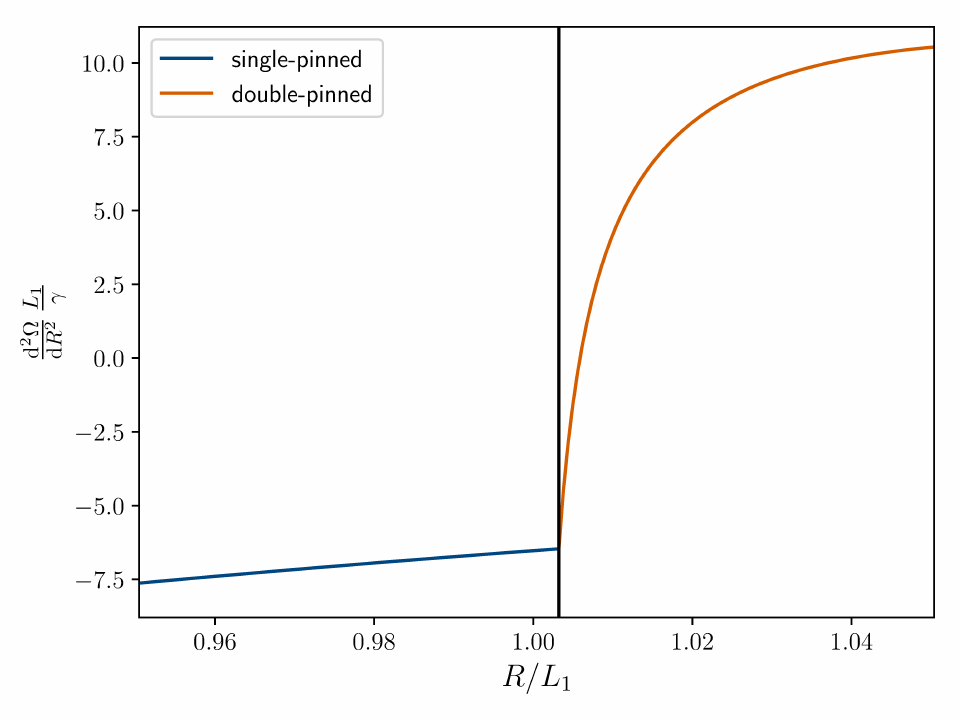}%
    \label{fig:ddomega_cw}%
  }%
  \\[0pt]%
  \subfloat{\columnwidth}{}{%
    \includegraphics[width=\textwidth]{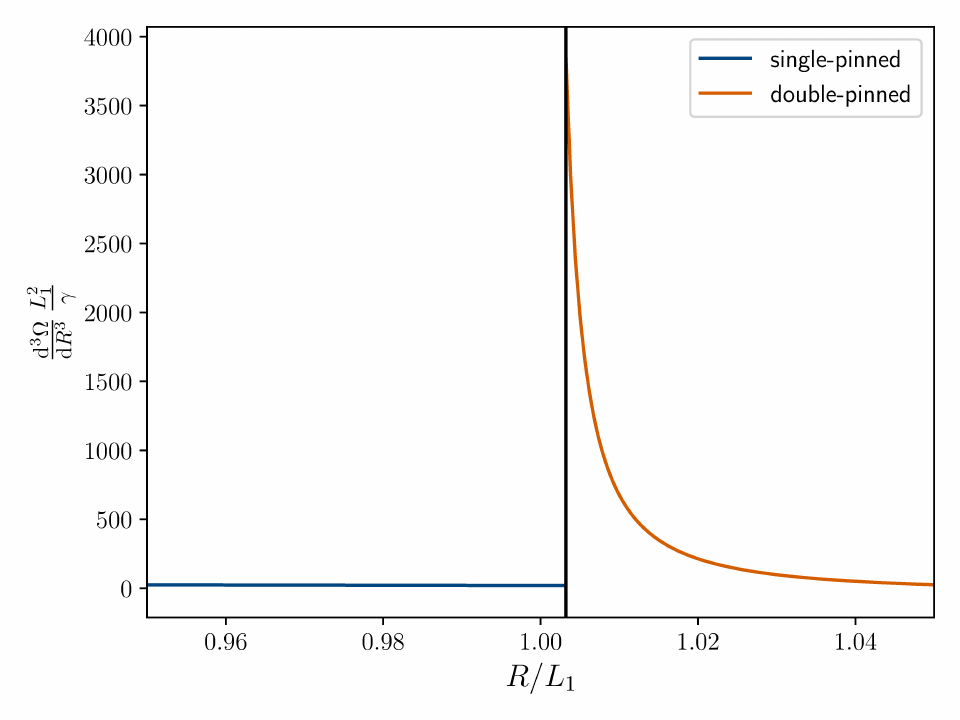}%
    \label{fig:dddomega_cw}%
  }%
  \caption{The illustrative behaviour of first (a), second (b) and third (c) derivatives of the grand potential of the single-pinned phase (for $R<R_{\rm depin}$) and the double-pinned phase (for
    $R>R_{\rm depin}$) where  $R_{\rm depin}\equiv\gamma/\delta p_{\rm depin}$ (denoted by the vertical line) corresponds to the depinning transition of the system formed of completely wet walls
    ($\theta=0$). The other parameters of the system are $L_2=2\,L_1$ and $\alpha=0.08$.}\label{domega_cw}
\end{figure}

The final remark regards the transition from a single- to  a double-pinned state in a fixed system geometry by varying the pressure. This \emph{depinning transition}, referring  to a meniscus
(de)pinning at the wider capillary end, occurs when $\theta_e'=\theta$, i.e., at the partial pressure
\bb
\delta p_{\rm depin}=\frac{2\gamma\cos(\theta+\alpha)}{L_2}\,, \label{depin}
\ee
as follows from Eq.~(\ref{dp2}). The free energy change (per unit length) associated with the transition is
\begin{widetext}
  \bb
  \Delta\Omega(R)\equiv\Omega_{\rm DP}^{\rm ex}-\Omega_{\rm SP}^{\rm ex}
  =\frac{\gamma}{4R}\left[L_2^2\cot\alpha+4R^2\left(\frac{\cos\theta\cos(\alpha+\theta)}{\sin\alpha}+\alpha+\theta-\cos^{-1}\left(\frac{L_2}{2R}\right)\right)-4L_2R\frac{\cos\theta}{\sin\alpha}+L_2\sqrt{4R^2-L_2^2}\right]\,.
  \label{depin2}
  \ee
\end{widetext}
It is straightforward to show that both $\Delta\Omega(R)=0$ and ${\rm d}\Delta\Omega(R)/{\rm d}R=0$ -- under the necessary condition that $\theta+\alpha<\pi/2$ -- for $R=R_{\rm depin}$, where $R_{\rm
      depin}\equiv\gamma/\delta p_{\rm depin}$ is the Laplace radius corresponding to the transition. For the second derivative of $\Delta\Omega(R)$ at the transition, one obtains
\bb
\left.\frac{{\rm d}^2 \Delta\Omega}{{\rm d}R^2}\right|_{R=R_{\rm depin}}=\frac{4\gamma\cos^3(\alpha+\theta)(\cot\alpha-\cot(\alpha+\theta))}{L_2}\,,
\ee
which vanishes \emph{only} for $\theta=0$. Finally, the third derivative of $\Delta\Omega(R)$ for completely wet walls at the depinning transition is
\bb
\left.\frac{{\rm d}^3 \Delta\Omega}{{\rm d}R^3}\right|_{R=R_{\rm depin}}=\frac{8\gamma\cos^2\alpha\cot^3\alpha}{L_2^2}\,,\;\;\;(\theta=0)\,,
\ee
which is already finite.

Hence, the change from a single- to a double-pinned state is accompanied by a continuous (depinning) phase transition, which (according to the classical Ehrenfest classification) is of second order
for systems formed of partially wet walls and of third order for completely wet walls. The behaviour of the excess grand potential near the transition is illustrated in Fig.~\ref{domega_pw} (for
partially wet walls) and in Fig.~\ref{domega_cw} (for completely wet walls).

\section{Summary and Concluding Remarks}

In this work, we studied condensation in capillaries formed of a pair of non-parallel walls, each of length $H$ and making an angle $\alpha$ with the vertical plane (say). We showed that the rotation
of the walls by the angle $\alpha$, breaking the reflection up-down symmetry, enriches the phase behaviour of the confined fluid substantially. The main features of the system phase behaviour are as
follows:

\begin{enumerate}

  \item The system exhibits capillary condensation only if the walls are sufficiently long, such that $H\ge L_1\sec\theta$ where $L_1$ is the capillary width at the narrow end and $\theta$ Young's
        contact angle of the walls.

  \item Furthermore, if the condition $H\ge L_1\sec\theta$ is obeyed, capillary condensation may occur only for systems with the opening angle $\alpha\le\alpha_{\rm max}$, where the marginal angle $\alpha_{\rm
            max}$ is given by the relation $\sin\alpha_{\rm max}=\cos\theta-L_1/H$. This generalizes the condensation condition for the semi-infinite system, $H\to\infty$,
        which is $\alpha_{\rm max}=\pi/2-\theta$, representing the wedge-wetting (filling) phase boundary.

  \item Capillary condensation is one of two types, termed single-pinning and double-pinning:

        \begin{enumerate}

          \item In the former case, the capillary is only partially filled with liquid, such that its lower meniscus is pinned at the narrow capillary end and makes edge contact angle $\theta_e$ with the walls as given
                by Eq.~(\ref{ke_sp2});
                the upper meniscus is inside the capillary and thus meets the walls with Young's contact angle $\theta$. The pressure, at which the system condenses to a single-pinned state, is given by the modified
                Kelvin equation (\ref{ke_sp}) in terms of $\theta_e$.

          \item If the condensation leads to a double-pinned state, the whole capillary is filled with liquid and the upper meniscus is pinned at the wider capillary end with a different edge contact angle
                $\theta_e'$. In this case, the condensation pressure is given by the modified Kelvin equation (\ref{ke_dp}), which can be expressed by either of the edge contact angles using  Eqs.~(\ref{dp1}),
                (\ref{dp2}), and (\ref{dp3}).

        \end{enumerate}

  \item The condensation scenario (when $H\ge L_1\sec\theta$) is as follows:

        \begin{enumerate}

          \item For $H>\tilde{H}(\approx4.6\,L_1)$, there exists a marginal value of Young's contact angle $\theta^+$, so that the condensation always leads to a double-pinned state for any $\alpha\le\alpha_{\rm
                    max}$. For $\theta<\theta^+$, the condensation may be of either type depending on $\alpha$, the variation of which leads to the re-entrant phenomenon:

                \begin{itemize}

                  \item For $\alpha<\alpha_1$, the condensation leads to a double-pinned state.

                  \item For $\alpha_1<\alpha<\alpha_2$, the condensation leads to a single-pinned state.

                  \item For $\alpha_2<\alpha<\alpha_{\rm max}$, the condensation leads to a double-pinned state again. As $\alpha$ tends to $\alpha_{\rm max}$, the menisci flatten as $R\sim(\alpha_{\rm max}-\alpha)^{-1}$.

                \end{itemize}

                The opening angles $\alpha_1$ and $\alpha_2$, for which single- and double-pinning coincide, are the solutions of Eq.~(\ref{triple}).

          \item In a narrow interval of $H$,  such that $H_c\le H\le\tilde{H}$ with $H_c\approx4.5\,L_1$, the interval of $\theta$ allowing for single-pinning is restricted by two marginal contact angles:
                $\theta^-<\theta<\theta^+$.

          \item For $H<H_c$, condensation always leads to a double-pinned state.

        \end{enumerate}

  \item Upon narrowing the slit, i.e., in the limit of $\alpha\to0$, the asymptotic properties of capillary condensation are as follows:

        \begin{enumerate}

          \item For semi-infinite walls, the condensation pressure, $p_{\rm SP}$,  tends to the one corresponding to capillary condensation in an infinite parallel slit, $p_{\rm CC}$, as $p_{\rm SP}-p_{\rm CC}\sim \alpha^{1/2}$ for
                partially wet walls, but substantially slowly,  $p_{\rm SP}-p_{\rm CC}\sim \alpha^{1/4}$, for completely wet walls. Analogous behaviour applies to the way the edge contact angle $\theta_e$
                approaches Young's contact angle $\theta$.

          \item For $H$ finite, the asymptotic behaviour is different because the system can now lower its free energy by condensing to a double-pinned state, so that the upper meniscus meets the walls
                with the edge contact angle $\theta_e'$ rather than $\theta$.
                Now, the condensation pressure approaches the one pertinent to a finite parallel slit linearly in $\alpha$---as does the difference in the edge contact angles, $\theta_e-\theta_e'$---both for
                partially and completely wet walls.

        \end{enumerate}

  \item For geometries allowing for single-pinned condensation, there is a continuous transition between a single-pinned and a double-pinned state by varying the pressure as given by
        Eq.~(\ref{depin}). The transition is of second order for partially wet walls and of third order for completely wet walls.

\end{enumerate}

We conclude with a few remarks concerning possible extensions of this study. Throughout this work we implicitly assumed that the capillary forces dominate the system behaviour and that gravity
effects can be neglected. This is the case when the pertinent capillary length is much larger than the characteristic lengths of the system (such as $L_1$). Allowing for gravity would make possible
to consider significantly larger ($\sim\mu m$) systems and supposedly would substantially enrich the system phase behaviour in view of the competing surface and gravity effects; this is
experimentally accessible using e.g. interferometric technique, as demonstrated by Moldover and Gammon for measuring elevation and adsorption thickness of $SF_6$ in wedge-like cavities
\cite{moldover}. On the other hand, employing more microscopic approaches that would allow to link the system behaviour with its molecular origin would also be desirable; this would, most notably, be
important for the case of $\theta=0$, when wetting layers adsorb at the walls effectively reducing the system size, something not accounted for within our analysis. Here, the direct link with real
experiments is less straightforward but possible using e.g. atomic force microscopy and we believe that a simple geometric modification of the recent experiments by Geim et al. \cite{geim} of
capillary condensation of water on mica or graphite ``walls'' of an atomic scale (few \aa ngst\"{o}ms) is feasible. More sophisticated treatments would be needed to capture further relevant aspects,
such as interfacial fluctuations, surface roughness or bulk criticality. Possible extensions also include modification of the system geometry by, for example, breaking the mirror symmetry, which can
further increase the number of condensation types. Some of these extensions will be subjects of our future work.

\begin{figure}[t]
  \includegraphics[width=\columnwidth]{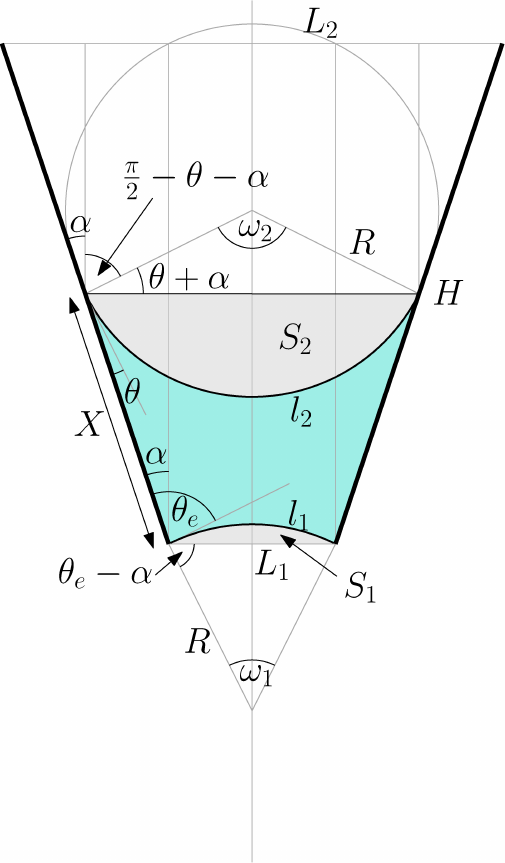}
  \caption{Detailed sketch of a single-pinned state. } \label{A1}
\end{figure}

\appendix

 \section{Derivation of formulae from Section II} \label{ap_a}

 Here, we present a detailed description of the geometric expressions that lead to the modified Kelvin equations for both single- and double-pinning condensations.

\subsection{Single-pinning condensation}

All the geometric measures that appear in Eq.~(4) for the excess grand potential can be determined easily using Fig.~\ref{A1} where all the important angles and distances are denoted. Specifically,
for the angles $\omega_1$ and $\omega_2$ we have that
 \bb
 \omega_1=\pi+2\alpha-2\theta_e \label{omega1}
 \ee
 and
 \bb
 \omega_2=\pi-2\alpha-2\theta\,,  \label{omega2}
 \ee
where $\theta$ is Young's contact angle at which the upper meniscus meets the walls and $\theta_e$ is the edge contact angle at which the lower meniscus meets the walls. From here, the expressions
(\ref{ell_left})  and (\ref{ell_right}) for $\ell_1$ and $\ell_2$ follow immediately. Furthermore, the shaded areas, $S_1$ and $S_2$, can now be determined as a difference between the areas of the
corresponding circular sectors, $R^2\omega_1/2$ and $R^2\omega_2/2$, and those of the isosceles triangles with the apex angle $\omega_1$ and $\omega_2$, respectively, leading to Eqs.~(9) and (10).
Finally, for the length $X$ of the walls which at contact with liquid, it holds
 \bb
  X=\frac{2R\cos(\alpha+\theta)-L1}{2\sin\alpha}\,.  \label{ap_X}
 \ee

After substituting for $\ell_1$, $\ell_2$ $S=S_{\rm tot}-S_1-S_2$, where the trapezoid area $S_{\rm tot}$ is given by Eq.~(8), and $X$ into the thermodynamic relation $\Omega_{\rm SP}^{\rm ex}=0$,
the condition (13), at which the single-pinning condensation occurs, is obtained. The second relation between the Laplace radius of curvature, $R$, and $\theta_e$ reads
 \bb
  R\cos(\theta_e-\alpha)=\frac{L_1}{2}\,,
 \ee
 as required by the system geometry.\\

\subsection{ Double-pinning condensation}

 \begin{figure}[t]
  \includegraphics[width=\columnwidth]{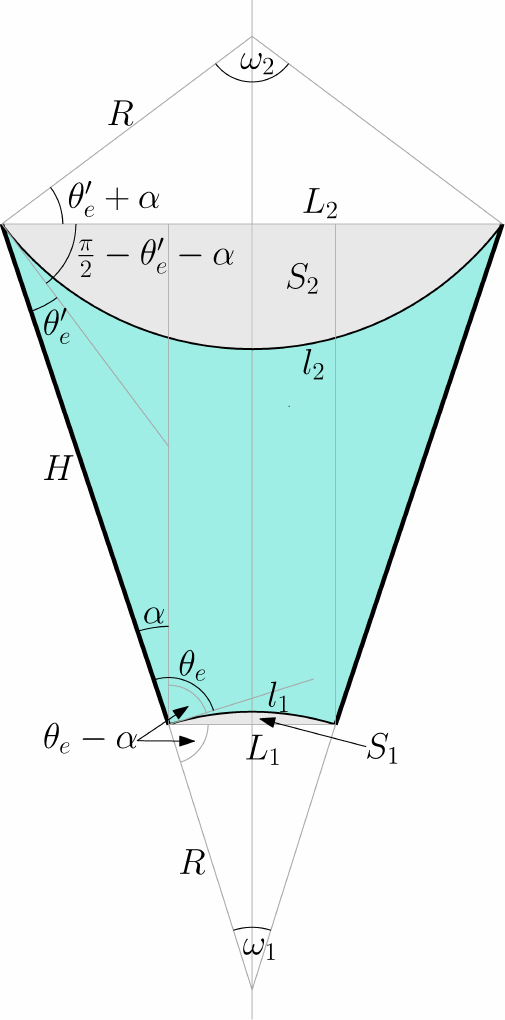}
  \caption{Detailed sketch of a double-pinned state. } \label{A2}
\end{figure}

Similar relations apply for the double-pinning condensation, which is illustrated in Fig.~\ref{A2}. While the expression (\ref{omega1}) for $\omega_1$ remains unchanged, the one for $\omega_2$ is now
given by the second edge contact angle, $\theta_e'$, as $\omega_2=\pi-2\alpha-2\theta_e'$. Hence, for the (shaded) areas of the circular segments we now have that
 \bb
 S_1=\frac{R^2}{2}\left(\pi-2\theta_e+2\alpha\right)-\frac{\sin(\theta_e-\alpha)RL_1}{2}
 \ee
 and
 \bb
 S_2=\frac{R^2}{2}\left(\pi-2\theta_e'-2\alpha\right)-\frac{\sin(\theta_e'+\alpha)RL_2}{2}\,,
 \ee
 which determines the area occupied by liquid $S=H\cos\alpha(L_1+L_2)/2-S_1-S_2$.

The phase boundary for the double-pinning condensation is given by the thermodynamic condition, $\Omega_{\rm DP}^{\rm ex}=0$, leading, after substituting for all the geometric measures, to Eq.~(20),
which is complemented by two geometric relations
 \bb
  R\cos(\theta_e-\alpha)=\frac{L_1}{2}
  \ee
  and
  \bb
   R\cos(\theta_e'+\alpha)=\frac{L_2}{2}\,,
   \ee
   forming now the set of three equations for the unknowns $R$, $\theta_e$ and $\theta_e'$.

\section{Derivation of Eq.~(\ref{alpha_max2})} \label{ap_b}

At the saturation, $\delta p=0$, both menisci in a double-pinned state must be flat and the simple geometry dictates that in this case $\theta_e-\alpha=\pi/2$ and $\theta_e'+\alpha=\pi/2$. The manner
at which the edge contact angles, $\theta_e$ and $\theta_e'$, acquire their limiting values follows by expanding of (\ref{dp1}) and (\ref{dp2}), which yields
 \bb
 \theta_e\to\frac{\pi}{2}+\alpha-\frac{L_1}{2R}\,,\;\;\;({\rm as}\;R\to\infty)
 \ee
 and
 \bb
 \theta_e'\to\frac{\pi}{2}-\alpha-\frac{L_2}{2R}\,,\;\;\;({\rm as}\;R\to\infty)\,,
 \ee
 implying that $\pi-\theta_e-\theta_e'\sim(L_1+L_2)/(2R)$, as $R\to\infty$. Dividing Eq.~(\ref{dp3}) by $R_{\rm DP}$ and considering the limit $R_{\rm DP}\to\infty$ one obtains that in this limit the
 condition for a double-pinning condensation reduces to
  \bb
  L_1+L_2+\frac{(L_1+L_2)\cos\theta}{\sin\alpha}=0\,,
  \ee
  which in terms of the parameter $H=(L_2-L_1)/(2\sin\alpha)$ can be expressed in the form of Eq.~(\ref{alpha_max2}).\\

\section{Derivation of Eq.~(\ref{triple})} \label{ap_c}

The change in the condensation type from single- to double-pinning occurs, when $X=H$ (cf. Fig.~1). In view of Eq.~(\ref{ap_X}) and the relation $H=(L_2-L_1)/2\sin\alpha$, the condition can be
written as
 \bb
 \cos(\theta_e-\alpha)=cq \label{ap_triple}
 \ee
 with $q\equiv L_1/L_2$ and $c\equiv\cos(\theta+\alpha)$. Using Eq.~(\ref{ap_triple}) and substituting for $\theta_e=\cos^{-1}(cq)+\alpha$ and $\sin(\theta_e-\alpha)=\sqrt{1-c^2q^2}$ into
 Eq.~(\ref{ke_sp2}) for $\theta_e$ at single-pinning condensation, Eq.~(\ref{triple}) is obtained.

\vspace{0.5cm}

\begin{acknowledgements}
  \noindent This work was financially supported by the Czech Science Foundation, Project No. 21-27338S. J.~J. acknowledges a financial support from the grant of Specific university research – grant No A1\_FCHI\_2023\_001.
\end{acknowledgements}


\begin{thebibliography}{99}
\bibitem{terminology}
    Here, we refer to finite slits, if their height (in the direction of the Cartesian $z$-coordinate, say) is finite; infinite, if the walls extend from $z=-\infty$ to $z=\infty$ (applicable, of course, only to parallel
    walls); and semi-infinite, if they span only the positive part of the $z$-axis.
\bibitem{semi_inf}
   Capillary condensation in a parallel semi-infinite slit occurs at the same pressure $p_{\rm CC}(L)$ as in the infinite case. However, a subtle difference is that whilts in the infinite slit the locations of the gas-like and
  liquid-like phases at the phase boundary are irrelevant, they do matter for the semi-infinite model. If the ambient phase is gas, the (``vertical'') portion of the gas-like phase is arbitrary but
  the one corresponding to the liquid-like phase must be infinite, which requires its location to be ``at the top'' of the capillary.
\end{thebibliography}
\end{document}